# A SIMPLE 2-APPROXIMATION ALGORITHM FOR MINIMUM MANHATTAN NETWORK PROBLEM

**[1]MD. MUSFIQUR RAHMAN SANIM, [2]SAFRUNNESA SAIRA, [3]FATIN FAIAZ AHSAN, [4]RAJON BARDHAN, [5]S.M. FERDOUS**

[1,2,3,4] STUDENT, AHSANULLAH UNIVERSITY OF SCIENCE & TECHNOLOGY, BANGLADESH
[5]ASSISTANT PROFESSOR, AHSANULLAH UNIVERSITY OF SCIENCE & TECHNOLOGY, BANGLADESH
E-mail: [1]sanim314159@gmail.com, [2]saira1108aust@gmail.com, [3]faiaz.ub@gmail.com, [4]rajonbardhan@gmail.com,
[5]ferdous.csebuet@gmail.com

**Abstract -** Given a n points in two dimensional space, a Manhattan Network G is a network that connects all n points with either horizontal or vertical edges, with the property that for any two point in G should be connected by a Manhattan path and distance between this two points is equal to Manhattan Distance. The Minimum Manhattan Network problem is to find a Manhattan network with minimum network length, i.e., summation of all line segment in network should be minimize. In this paper, we proposed a 2-approximation algorithm with time complexity O(|E|lgN) where |E| is the number of edges and N is the number of nodes. Using randomly generated datasets, we compare our result with the optimal one.

## I. INTRODUCTION

Let any two points p,q in two dimensional space and these two points are connected by a rectilinear path, a path that contains only horizontal or vertical line segments as edges. Manhattan path is same as a rectilinear path with length exactly as the Manhattan distance between these two points. Formally, the Manhattan distance between point's p and q is,

$$distance(p,q) := |p.x - q.x| + |p.y - q.y|.$$

where $p.x$ and $p.y$ denote the $x$ and $y$ coordinate of point $p$.

In this problem, we have set of $n$ points, T in two dimensional space and we need to construct a Manhattan network G on T such that any two point from T should have a Manhattan path containing the line segments or edges from G. Minimum Manhattan Network (MMN) Problem is to construct a Manhattan network G on T having minimum network length.
MMN has various applications in geometric network design. Most of the chips contain circuit path as rectilinear path. It has also applications in distributed algorithms, city planning and network layout [3].
The rest of the paper is organized as follows. In section II we will discuss the related works regarding MMN. Section III will be dedicated for the proposed algorithm. In section IV, we will analyze the run time and approximation ratio of the proposed algorithm. Before concluding in section VI we will list some preliminary experimental results in section V.

## II. LITERATURE REVIEW

MMN has applications in circuit building, city planning and computational biology. Due to its huge application, this problem attracted researchers. The MMN problem is formally introduced by Gudmundsson et al. In that paper he showed an $O(n^3)$ algorithm with approximation ratio 4. He also showed a 8-approximation algorithm that runs faster ($O(nlogn)$ time). He also argued about the NP-hardness of the problem and whether there are 2-approximation algorithm existed or not for MMN[6].
An $O(n^3)$ time 2-approximation algorithm was constructed in a distinct method by Kato et al.. The main concept is to infer if the graph is a Manhattan Network or not. A primitive approach is assuring if Manhattan path prevailed for all $O(n^2)$ pairs of nodes, but they proved that only O(n) specific pair of points are sufficient [9]. Benkert et al. [1] fabricated a 3-approximation algorithm applying almost identical idea as Seibert et al.[11] which requires linear space and runs in O(nlgn) time [1].
Benkert et al. [2] also suggested a mixed integer programming formulation. Chepoi et al. [3] generated a rounding 2-approximation algorithm founded on an LP-formulation of the MMN problem by applying the concept of Pareto front and strip-staircase decomposition [4]. Afterwards, basing on the notion of dynamic programming Guo et al. developed a 2-approximation algorithm. This solution runs within $O(n^2)$ time [7]. Associating with a simple greedy strategy they also submitted another approximation algorithm having analogous approximation ratio and a better running time of O(nlgn) [8]. Seibert et al. [11] came up with a 1.5-approximation algorithm. Nevertheless their proof may be incorrect showed by Chepoi et al. [3].
To that extent the complexity of this problem remained unknown. By reducing 3 –SAT to this problem Chin et al. [4] exhibited that Minimum Manhattan Network in 2 dimensions is strongly NP-complete. Munoz et al. [10] proved this problem to be NP-hard in 3 dimensions. They implied a 3-approximation algorithm which is the first approximation algorithm in 3 dimensions for a restricted version of this problem [10].

First approximation algorithm for Generalized Minimum Manhattan Network was conjured up by Das et al. [5]. They came up with an algorithm with an approximation ratio of $O(\lg^{d+1} n)$ for an arbitrary dimension d [5].

## III. PROPOSED ALGORITHM

The input of our algorithm is a set of points in 2D. We termed these points as "main" points. We also select a few more points on the 2D plane using a divide and conquer approach. These extra points are termed as "demo" points. We then formulate a graph using these two kinds of points as nodes and rectilinear paths between these points as edges. We then compute Minimum Spanning Tree (MST) of this graph. We run a post processing step on the tree to eliminate the unnecessary edges for MMN. The detailed algorithm is presented in Algorithm 1.

### A. Graph Construction

#### a) Node Construction:

Let S be the set of n points in 2D space (main points). We term this set to be the main set. Let also X and Y denote the x-coordiantes and y-coordinates of all the points in S. This main set is divided into four quadrants at point $c(x_m, y_m)$, where c is the median of the main set S. More formally,

$$c = \left( x_m = median(X), y_m = median(Y) \right).$$

For each point $p(x, y)$ in the set S, we introduce (at most) two new points $p_1(x1, y1)$ and $p_2(x2, y2)$ where, $x_1 = x_m, y_1 = y, x_2 = x, y_2 = y_m$. These new sets of points will be termed as demo points. Note that if any demo point matches with a main point, we will ignore that demo point and retain the main point. Let D denotes the set that holds the demo points. We then apply same procedure recursively on the four quadrants computed earlier by the median c. In figure 1, we show the main and demo nodes by red and blue color. The details algorithm is shown in algorithm 2.The nodes of the graph are,

$S \cup D.$

#### b) Edge Construction:

We choose rectilinear edges between the points in S. The edges are constructed by two phases. Firstly we proceed through a horizontal scan on the set of points. We divide the set of points into a several group based on their y coordinate. Each group consists of the points with equal y coordinate value. Within each group we sort the points in ascending order by their x-coordinate value. Let $a1, a2, \ldots, al$ be the sorted points in a group. We add horizontal undirected edges from $a_i$ to $a_{i+1}$, where i = 1 to l − 1. The weight of each edge is $a_i.x - a_{i+1}.x$.

Similar to first phase we run a vertical scan on the original set of points in S. We divide those points into several groups by the equality of their x-coordinate value and. We sort the points in a particular group by their y-coordinate value. Similar to the previous phase, weadd vertical weighted edge with weight equals the difference between their y-coordinate values. The detailed pseudo code of EdgeConstruction is shown in Algorithm 3.A sample graph is shown in Figure 1.

### B. Manhattan Network Construction

The graph constructed from G has all rectilinear edges. To find a manhattan network on the graph G, it is sufficient to find a spanning tree on G. We will use the well-known Minimum spanning algorithms [13] for computing MST. The MST contains some extra edges due to the demo nodes. So we need a post processing steps to remove to unnecessary edges. We will traverse the graph using graph search algorithm (such as DFS or BFS) [12]. When we find a node who is a demo node and all of its neighbors are also demo nodes we will remove the sub graph induced by these nodes.

So, the edge between two demo nodes will be eliminated if the edge is not necessary to reach any main node. The feasible edges are shown in Figure 1. The detailed pseudo code of Edge Remove is shown in Algorithm 4.Final Manhattan Network is shown in Figure 2.

---

**Algorithm 1 MMNFA (Minimum Manhattan Network Fixed Algorithm)**

---

Take global variable set of nodes S
Take global variable of graphmmnPath
Take empty Graph G,T
mainNodes←Read dataset
S←append(S,mainNodes)
TakingDemoNodes(mainNodes)
G←EdgeConstruction(S)
T←MST(G)
DFS(any node from mainNodes)
mmnCost← summation of all edges weight from mmnGraph

---

**Algorithm 2 TakingDemoNodes**

---

**function**TakingDemoNodes(Nodes)
**if**Nodes.length<= 1:
return
**endif**
Take node P
P.x←median of all X-coordinate of Nodes
P.y←median of all Y-coordinate of Nodes

S←append(S,P)
**for**each node(x,y) fromNodes

S←append(S, node(X,P.y))
S←append(S, node(P.x,Y))
` **end for**
Take four empty vectors A,B,C,D
**for**each node(x,y) from Nodes
**if** x<P.x and y<P.y
A← append(A, node(x,y))
**else if** x<P.x and y>P.y
B← append(B, node(x,y))
**else if** x>P.x and y<P.y
C← append(C, node(x,y))
**else if** x>P.x and y>P.y
D← append(D, node(x,y))
**end if**
**end for**
TakingDemoNodes(A)
TakingDemoNodes(B)
TakingDemoNodes(C)
TakingDemoNodes(D)
**end function**

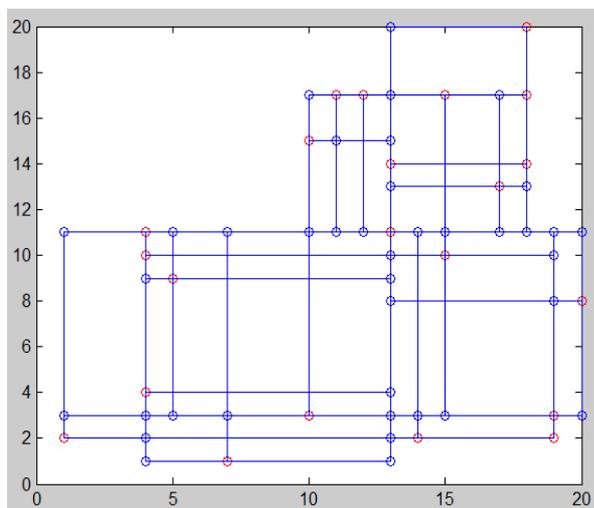

**Figure 1:** After graph construction, red circles are main nodes and blue circles are demo nodes.

---

**Algorithm 3 EdgeConstruction**

---

**function**EdgeConstruction (S)
Take Graph G
**for**each pair of node(x,y) from S
Take a new edge E
E.firstNode←node(x,y)
node($x_1$,y) ←from node(x,y) find the nearest left side node from S
**if**node($x_1$,y) ≠ null
E.secondNode←node($x_1$,y)
E.weight←|$x_1$-x|
G←append(G,E)
**end if**
node(x,$y_1$) ←from node(x,y) find the nearest upper node from S
**if**node(x,$y_1$) ≠ null

E.secondNode←node(x,$y_1$)
E.weight←|$y_1$-y|
G←append(G,E)
**end if**
**end for**
**return** G
**end function**

---

**Algorithm 4 EdgeRemove(nNode)**

---

**function**EdgeRemove(nNode)
nNode is visited
flag←FALSE
**for**each edge E contain nNode
node(x,y) ←adjacent node from nNode in this edge
**if**node(x,y) is not visited
f←EdgeRemove(node(x,y))
**if**f is TRUE
mnnPath←append(mmnPath,E)
**end if**
flag←flag or f
**end if**
**end for**
**if**nNode is in mainNodes
flag←TRUE
**end if**
**return**flag
**end function**

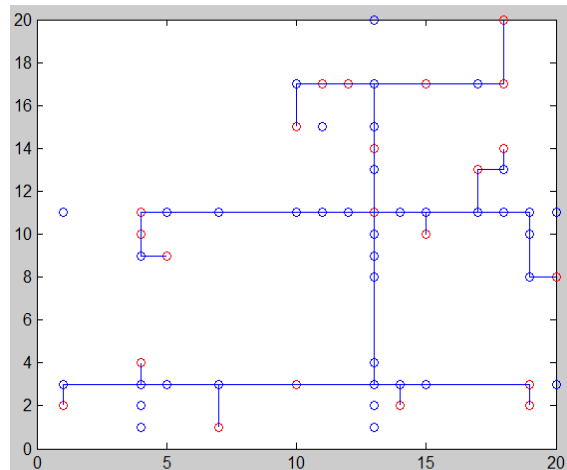

**Figure 2: Final Manhattan Network.**

## IV. ANALYSIS

### A. Calculating maximum number of nodes

Assume that number of main nodes is n. In TakingDemoNode, after each iteration nodes are divided into four parts. So the depth of the recursion of TakingDemoNodes,
d =log₄(n).
In each recursion step, for each node, we take two extra nodes as demo nodes. So the number of nodes after adding demo nodes are at most
N=n+2*n*d.

### B. Calculating Maximum Number of edges

There are at most four edges from each nodes in S U D. So, the number of edges in graph G is at most,

|E| = 4*N/2 = 2*N

### C. Time complexity analysis

• **TakingDemoNodes**:From IV(A) we saw that depth of recursion for TakingDemoNodes is d That's mean a main node can appeared at most d times in TakingDemoNodes. So time complexity is O(n*d) or O(n * log(n)).

• **EdgeConstruction**:In here we take a node and construct an edge with nearest node in O(1).We do it for all nodes (main and demo). So time complexity is O(N).

• **Minimum Spanning Tree(MST)**:We use Kruskal algorithm which time complexity is O(Elog(E)). [13]

• **EdgeRemove**:We use DFS algorithm for this which time complexity is O(N + E).

So finally the time complexity of our proposed algorithm MMNFA is **O(|E|log(E)).**

### D. Approximation analysis

Assume that we have a network where A, B, C are main nodes and X, Y are demo nodes (Figure 3).A and B are connected. Now we want to connect C with this network. Let the distance between C to X and C to Y are $D_1$ and $D_2$ respectively. There might be three cases which are discussed as follows:

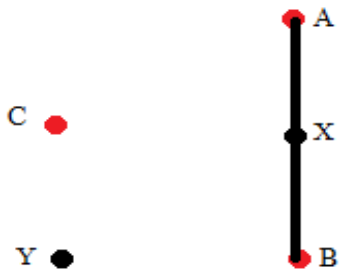

**Figure 3: A network of Main nodes (Red dots) and Demo nodes (Black dots)**

#### a) Case1 ($D_1 < D_2$)

In this case, MST algorithm will pick$D_1$ first. Then it will take either$C$ to$Y$ or $B$ to$Y$. But after running the$EdgeRemove$function that edge will be eliminated (Figure 4).

This actually gives us optimal answer. So, the optimal network length for connecting C,

L=$D_1$

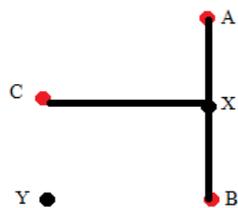

**Figure 4: Scenario 1 (where $D_1 < D_2$).**

#### b) Case2 ($D_1 = D_2$)

In this case MST algorithm will take either$D_1$ or $D_2$. In the worst case at first MST will take $D_2$ then take the edge between Y and B (Figure 5).

Network Length for connecting C,

L =$D_2$ + $D_1$

=$D_1$ + $D_1$

=2 * $D_1$

In this case L is 2 times greater than optimal.

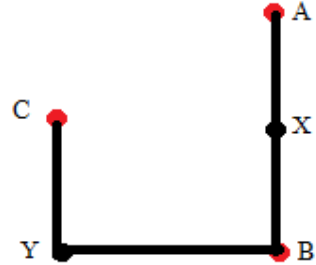

**Figure 5: Scenario 2 (Where $D_1=D_2$) and Scenario 3 (Where $D_1>D_2$).**

#### c) Case3 ($D_1>D_2$)

In this case at first MST algorithm will$pick D_2$. Then in the worst case it will take edge between B to Y (Figure 5).

Network Length for connecting C,

L = $D_2$ + $D_1$

which is between D1 and 2*$D_1$.

### EXPERIMENTAL RESULT

To our best knowledge, there are no any benchmark dataset for MMN(Minimum Manhattan Network) [10]. Here we introduced a random set of data which contains 10 test cases.

The tableI shows the number of nodes, total nodes after taking demo nodes, total number of edges and network length of MMNFA after running our proposed algorithm.

**TABLE I. Result of the random dataset**

| Number of main nodes | Total number of nodes (main nodes with demo node) | Total number of edges | Net. Length in MMNFA |
|---|---|---|---|
| 1000 | 7408 | 13724 | 29062 |
| 1000 | 7372 | 13653 | 30114 |
| 1000 | 7320 | 13557 | 29075 |
| 1000 | 7395 | 13696 | 29510 |
| 1000 | 7278 | 13473 | 28867 |
| 1000 | 7377 | 13645 | 29352 |
| 1000 | 7442 | 13783 | 29705 |
| 1000 | 7297 | 13502 | 28738 |
| 1000 | 7397 | 13687 | 29496 |
| 1000 | 7213 | 13355 | 28416 |

The table II shows the difference in network length between our proposed MMNFA algorithm and optimal network length where all the test cases contain 5 to 8 nodes. As the optimal algorithm exponential in time, we only could experiment with few (5 to 8 in our case) nodes.

TABLE II. Comparison BETWEEN MMNFA and optimal MMN.

| Network length in MMNFA | Optimal network length | Difference |
|---|---|---|
| 15 | 15 | 0 |
| 21 | 17 | 4 |
| 15 | 15 | 0 |
| 15 | 15 | 0 |
| 14 | 14 | 0 |
| 16 | 15 | 1 |
| 18 | 18 | 0 |
| 14 | 13 | 1 |
| 15 | 15 | 0 |

## CONCLUSION

In this paper we developed a simple algorithm to find 2- approximate manhattan network of a set of points. The algorithm is easy to implement and efficient. Further experimentation, we try to reduce the demo nodes and edges. Because modern chips contain millions of circuit paths, for this we need to reduce the edges for efficient circuit design.

```
#include<bits/stdc++.h>
usingnamespacestd;
struct node {
int x, y, id;
node () {}
node(int _x, int _y, int _id) {
x = _x;
y = _y;
id = _id;
}
booloperator< (const node &a)const {
if(this->x !=a.x) returnthis->x <a.x ;
returnthis->y <a.y;
}
};
```

```cpp
struct edge {
node u;
node v;
int wh;
edge () { }
edge(node _u, node _v) {
u = _u;
v = _v;
wh= abs(u.x-v.x) + abs(u.y-v.y);
}
bool operator< (const edge &a)const {
return this->wh<a.wh;
}
};
const int mx =1e5+7;
set<node> S;
vector<edge>mmnPath;
int mmnNetworkLength;
vector<node>mainNodes;
node up[mx], Hash[mx];
int prv[mx];
bool vis[mx];
vector<int>adj[mx];
int cntNode;
void init() {
S.clear();
mmnPath.clear();
for(int i=0; i< mx; ++i) up[i].id =-1;
memset(vis, false, sizeof vis);
mmnNetworkLength=0;
cntNode=0;
}
void read() {
int n, x, y;
cin>> n;
for(int i=0; i< n; ++i) {
cin>> x >> y;
mainNodes.push_back({x,y,0});
}
}
void TakingDemoNodes(vector<node>_mainNodes){
if(_mainNodes.size() <=1) return;
int n = _mainNodes.size();
vector<int>X, Y;
for(auto p : _mainNodes){
X.push_back(p.x);
Y.push_back(p.y);
}
sort(X.begin(), X.end());
sort(Y.begin(), Y.end());
node P = node (X[n/2], Y[n/2], 0);
vector<node >A,B,C,D;
for (auto p : _mainNodes){
S.insert(node(P.x, p.y,0));
S.insert(node(p.x, P.y,0));
if(p.x<P.x&&p.y<P.y) {
A.push_back(p);
} else if(p.x<P.x&&p.y>P.y) {
B.push_back(p);
} else if(p.x>P.x&&p.y<P.y) {
```

```
C.push_back(p);
} elseif(p.x>P.x&&p.y>P.y) {
D.push_back(p);
}
}
TakingDemoNodes(A);
TakingDemoNodes(B);
TakingDemoNodes(C);
TakingDemoNodes(D);
}
vector<edge>EdgeConstruction(set<node>_S) {
vector<edge>G;
node lst= node(-1,-1,0);
for(auto p : _S) {
node u = p;
u.id =cntNode;
Hash[cntNode] = u;
if(lst.x==u.x) {
G.push_back({u, lst});
}
if(up[u.y].id !=-1) {
G.push_back({u, up[u.y]});
}
up[u.y] = u;
lst= u;
++cntNode;
}
return G;
}
int Parent(inti){
if(i==prv[i])returni;
return (prv[i]=Parent(prv[i]));
}
intisUnion(inta,int b){
return Parent(a)==Parent(b);
}
voidmakeUnion(inta,int b ){
prv[Parent(a)] = Parent(b);
}
vector<edge>MST(vector<edge>G) {
sort(G.begin(), G.end());
vector<edge>T;
for(inti=0; i< mx; ++i) prv[i] =i;
int x =0;
for(auto e : G) {
if(!isUnion(e.u.id, e.v.id)) {
makeUnion(e.u.id, e.v.id);
T.push_back(e);
adj[e.u.id].push_back(e.v.id);
adj[e.v.id].push_back(e.u.id);
++x;
}
}
return T;
}
boolEdgeRemove(int u){
vis[u] =true;
bool flag =false;
for(autonxt:adj[u]){
if(!vis[nxt]) {
```

```
bool f =EdgeRemove(nxt);
edge e = {Hash[u], Hash[nxt]};
if(f){
mmnPath.push_back(e);
mmnNetworkLength+=e.wh;
}
flag |= f;
}
}
node p = Hash[u];
p.id =0;
if(binary_search(mainNodes.begin(), mainNodes.end(), p)) {
returntrue;
}
return flag;
}
void MMNFA() {
init();
read();
sort(mainNodes.begin(), mainNodes.end());
S.insert(mainNodes.begin(), mainNodes.end());
TakingDemoNodes(mainNodes);
vector<edge>G =EdgeConstruction(S);
vector<edge>T = MST(G);
for(inti=0; i<cntNode; ++i){
node u = Hash[i];
u.id =0;
if(binary_search(mainNodes.begin(), mainNodes.end(), u)) {
EdgeRemove(i);
break;
}
}
}
int main() {
MMNFA();
}
```

★ ★ ★